\documentclass[10pt,twocolumn,pre,byrevtex,showpacs,superscriptaddress,showkeys,bibnotes]{revtex4}
\usepackage{graphics}

\begin{document}

\noindent\parbox{\textwidth}{\small{
Proceedings of the 3rd International Conference on\\
News, Expectations and Trends in Statistical Physics (NEXT-$\Sigma\Phi$ 2005)\\
13-18 August 2005, Kolymbari, Crete, Greece}}
\bigskip

\title{Metastability within the generalized canonical ensemble}

\author{H. Touchette}
\email{htouchet@alum.mit.edu}
\affiliation{School of Mathematical Sciences, Queen Mary, University of London,
London E1 4NS, UK}

\author{M. Costeniuc}
\affiliation{Department of Mathematics and Statistics, University of Massachusetts,
Amherst, MA 01003, USA }

\author{R. S. Ellis}
\email{rsellis@math.umass.edu}
\affiliation{Department of Mathematics and Statistics, University of Massachusetts,
Amherst, MA 01003, USA }

\author{B. Turkington}
\email{turk@math.umass.edu}
\affiliation{Department of Mathematics and Statistics, University of Massachusetts,
Amherst, MA 01003, USA }

\date{\today}

\begin{abstract}
We discuss a property of our recently introduced generalized canonical
ensemble (J.\ Stat.\ Phys.\ 119 (2005) 1283). We show that this ensemble can
be used to transform metastable or unstable (nonequilibrium) states of the
standard canonical ensemble into stable (equilibrium) states within the
generalized canonical ensemble. Equilibrium calculations within the
generalized canonical ensemble can thus be used to obtain information about
nonequilibrium states in the canonical ensemble.
\end{abstract}

\pacs{05.20.Gg, 65.40.Gr, 64.60.My}

\keywords{Nonequivalent ensembles, nonconcave entropy, generalized canonical
ensemble, metastability}

\maketitle

The calculation of the equilibrium properties or \textit{states} of systems
having a nonconcave microcanonical entropy function (entropy as a function
of the mean energy) is plagued by a fundamental problem. Because of the
nonconcavity of the entropy function, these systems possess energy-dependent
equilibrium states in the microcanonical ensemble that have no equivalent
whatsoever within the set of temperature-dependent equilibrium states of the
canonical ensemble \cite{eyink1993,gross1997,gross2001,ellis2000,draw2002,touchette2004}.
These so-called \textit{nonequivalent microcanonical states} cannot, as a result,
be calculated or predicted from the point of view of the canonical ensemble.
To assess them, one must either resort to the microcanonical ensemble, with
all the analytical or numerical complications that this may imply, or else
one must find a method with which one can study the nonequilibrium (i.e.,
metastable or unstable) states of the canonical ensemble. Indeed, it is
known that nonequivalent microcanonical states correspond in general to
metastable or unstable critical states of the canonical ensemble
\cite{eyink1993,antoni2002,touchette2003,campa2004,touchette2005}. Hence, if one
is able to calculate all the critical states of the canonical ensemble
(equilibrium and nonequilibrium), then one has a complete knowledge of all
the microcanonical equilibrium states, including the nonequivalent ones.

Recently, we have followed this last line of thought to propose a
generalization of the canonical ensemble which effectively transforms some
of the metastable and unstable critical states of the canonical ensemble
into equilibrium states of a generalized canonical ensemble
\cite{costeniuc22005,costeniuc2006}. The generalized canonical ensemble is
defined by the following probability measure
\begin{equation}
P_{g,\alpha }(\omega )=\frac 1{Z_g(\beta )}e^{-\alpha H(\omega )-ng(H(\omega
)/n)},  \label{gencan1}
\end{equation}
where
\begin{equation}
Z_g(\alpha )=\int e^{-\alpha H(\omega )-ng(H(\omega )/n)}d\omega
\label{gpf1}
\end{equation}
represents a generalized canonical partition function. In these expressions,
$\omega $ denotes the microstate of an $n$-body system with Hamiltonian $
H(\omega )$, $\alpha $ is an analog of an inverse temperature, and $g$ is
some continuous function of the mean energy $H(\omega )/n$ which is as yet
unspecified. The canonical ensemble is a special instance of this measure
obtained, obviously, by choosing $g=0$ and $\alpha =\beta $. The case of
quadratic functions $g(u)=u^2$ defines a statistical-mechanical ensemble
known as the Gaussian ensemble \cite{challa21988,challa1988}.

We have reported many properties of the generalized canonical ensemble in
two recent papers \cite{costeniuc22005,costeniuc2006}---one containing
rigorous results and their detailed proofs \cite{costeniuc22005}, and
another one which is more ``physical'' in its presentation
\cite{costeniuc2006}. The main point formulated in these papers is that the
generalized canonical ensemble can be used, with suitable choices of the
function $g$, to assess all the microcanonical equilibrium states of a
system, including those not found at equilibrium in the canonical ensemble.
In other words, for suitable choices of $g$, the generalized canonical
ensemble can be made equivalent with the microcanonical ensemble in the
thermodynamic limit. The conditions on $g$ that ensure equivalence are
stated in these papers both in terms of the microcanonical entropy function
and in terms of a generalized canonical free energy function defined from
$Z_g(\alpha )$.

Our aim in this short contribution is to discuss one interesting property of
the generalized canonical ensemble that we have alluded to above, namely
that the generalized canonical ensemble can be used to transform metastable
or unstable (nonequilibrium) states of the canonical ensemble into stable
(equilibrium) states. We shall try to explain herein how this is possible
and how this works in practice.

The framework of our discussion is the same as in our previous papers
\cite{costeniuc22005,costeniuc2006}. We consider an $n$-body system described by
some Hamiltonian function $H(\omega )$, with $\omega $ denoting the
microstate of the system. In the canonical ensemble, that is, in a situation
where the system is in contact with a fixed-temperature heat bath, all the
equilibrium properties of the system are calculated using Gibbs's canonical
measure
\begin{equation}
P_\beta (\omega )=\frac{e^{-\beta H(\omega )}}{Z(\beta )},\qquad Z(\beta
)=\int e^{-\beta H(\omega )}d\omega .
\end{equation}
To calculate, for example, the equilibrium or thermody-namic-limit value of
the mean Hamiltonian $h(\omega )=H(\omega )/n$ associated with a given value
$\beta $ of the inverse temperature, one first writes the probability
measure for $h$, i.e.,
\begin{equation}
P_\beta (u)=\int_{\{\omega :h(\omega )=u\}}P_\beta (\omega )d\omega =\int
\delta (h(\omega )-u)P_\beta (\omega )d\omega ,
\end{equation}
and then finds the global maximum of $P_\beta (u)$ in the limit where
$n\rightarrow \infty $. The result of these steps is well-known: the global
maximum of $P_\beta (u)$, which corresponds again to the equilibrium value
of $h(\omega )$ in the canonical ensemble with inverse temperature $\beta $,
is given by the global minimum of the function
\begin{equation}
F_\beta (u)=\beta u-s(u)
\end{equation}
evaluated at fixed $\beta $. The function $s(u)$ in this expression is the
microcanonical entropy function defined by the usual limit
\begin{equation}
s(u)=\lim_{n\rightarrow \infty }\frac 1n\log \int \delta (h(\omega
)-u)d\omega .  \label{ms1}
\end{equation}
In the remainder, we shall denote the global minimum of $F_\beta (u)$ at
fixed $\beta $ by $u_\beta $.

\begin{figure*}[t]
\resizebox{5.5in}{!}{\includegraphics{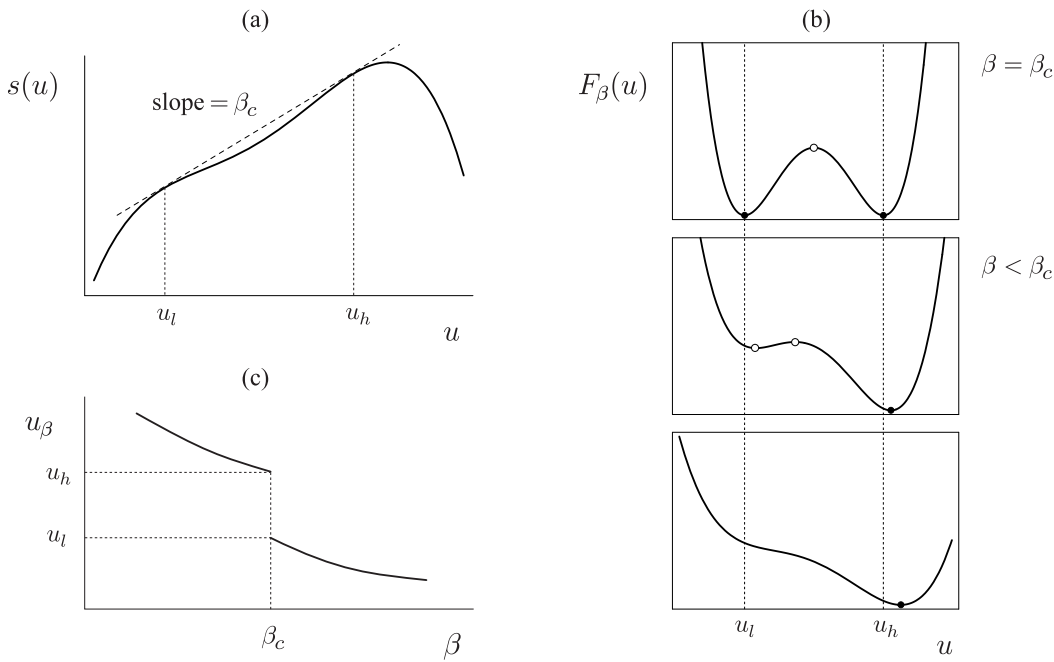}}
\caption{(a) Generic nonconcave microcanonical entropy function.
(b) The function $F_\protect\beta (u)$ constructed from $s(u)$ shows,
for some values of $\protect\beta$, a local minimum
and a local maximum in addition to a global minimum. The global minimum
defines the canonical equilibrium mean energy $u_\protect\beta$, while the
local min and local max correspond to metastable and unstable mean
energies of the canonical ensemble, respectively.
(c) Behavior of $u_\protect\beta$ as a function of $\protect\beta$.
The nonconcavity of $s(u)$ leads to a first-order phase transition in the canonical
ensemble. This is reflected in $F_\protect\beta (u)$ by the fact that the
global minima of this function never enter the range $(u_l,u_h)$.}
\label{metae1}
\end{figure*}

This calculation of the equilibrium value $u_\beta $ enables us to
understand in a very simple manner how metastable and unstable states can
arise in the canonical ensemble. Also, we can understand with it why these
states must arise in connection with nonequivalent ensembles and nonconcave
entropies. The key points to observe here are the following (see also
Fig.~\ref{metae1}):

(i) There can be mean-energy values $u$ in the domain of $F_\beta (u)$
that do not correspond to global minima of $F_\beta (u)$ for any $\beta $.
In other words, the set of all possible values of $u_\beta $, i.e., its
range, can be a proper subset of the domain of $F_\beta (u)$, which is the
same as the domain of $s(u)$. This happens when there is a first-order phase
transition in the canonical ensemble; see Fig.~\ref{metae1}.

(ii) The domain of $s(u)$ indicates which values of the mean energy are
accessible to the microcanonical ensemble, while the range of $u_\beta $
indicates which values of the mean energy are accessible to the canonical
ensemble by varying $\beta $. The previous point, therefore, implies the
nonequivalence of the microcanonical and canonical ensemble---there are
energy values accessible to the microcanonical ensemble, but not to the
canonical ensemble.

(iii) Those values of $u$ that are not global minima of $F_\beta (u)$ for
any $\beta $ can be local minima or local maxima of $F_\beta (u)$ for
certain values of $\beta $. The local minima correspond physically to
metastable states of the canonical ensemble, while the local maxima
correspond to unstable states. Note that the critical points of $F_\beta (u)$
(minima, maxima or saddle-points) are given by $F_\beta ^{\prime }(u)=0$ or,
equivalently, $s^{\prime }(u)=\beta $ if $s(u)$ is differentiable.

(iv) In order to have local minima and local maxima of $F_\beta (u)$,
$F_\beta (u)$ must be nonconvex, which implies that $s(u)$ must be
nonconcave. Accordingly, the nonconcavity of $s(u)$ is a necessary condition
for having metastable and unstable states in the canonical ensemble (it is
also a sufficient one).

These points have all been discussed at various levels of rigor in the
literature; see, e.g., Ref.~\cite{gunton1983} for a list of references on
the subject of metastable and unstable states, and Ref.~\cite{touchette2005}
for more recent references on the connection with nonequivalent ensembles.
For our purpose, what is important to keep in mind is that, for a given
$\beta $, $F_\beta (u)$ may have local minima and local maxima in addition to
one or more global minima corresponding to equilibrium states, and that all
of these critical points of $F_\beta (u)$ satisfy the equation $s^{\prime
}(u)=\beta $ if $s(u)$ is differentiable. Additionally, the stability of
these critical points is determined by the sign of $s^{\prime \prime }(u)$,
provided that $s(u)$ is twice differentiable. Indeed, by basic calculus, we
have that if a point $\hat{u}$ satisfying $F_\beta ^{\prime }(\hat{u})=0$ is
such that $F_\beta ^{\prime \prime }(\hat{u})=-s^{\prime \prime }(\hat{u})>0$,
then that point must be a minimum of $F_\beta (u)$, either local of
global. If the same point is such that $F_\beta ^{\prime \prime }(\hat{u}
)=-s^{\prime \prime }(\hat{u})<0$, then it is a local maximum of $F_\beta
(u) $.

Now, let us consider in more details the situation in which $\hat{u}$ is a
microcanonically-allowed but canonically-unallowed value of $h(\omega )$ at
equilibrium; that is, suppose that $\hat{u}$ is in the domain of $s(u)$ but
not in the range of $u_\beta $. As we have noted above, $\hat{u}$ can
correspond to a metastable or unstable critical point of $F_\beta (u)$, but
never to a stable point of $F_\beta (u)$, for the simple reason again that
the stable points of $F_\beta (u)$ correspond to the equilibrium mean
energies $u_\beta $ in the canonical ensemble. However, and this is the
central point of this paper, \textit{$\hat{u}$ can correspond to an
equilibrium mean energy of the generalized canonical ensemble defined by
Eq.(\ref{gencan1}).} The analog of $F_\beta (u)$ in the generalized canonical
ensemble is the function
\begin{equation}
F_{g,\alpha }(u)=\alpha u+g(u)-s(u).
\end{equation}
Therefore, the previous claim can be rephrased alternatively as following:
\textit{there is a choice of function $g$ and value $\alpha $
that make $\hat{u}$ a global minimum of $F_{g,\alpha }(u)$.}

To demonstrate our claim, let us study the critical points of $F_{g,\alpha
}(u)$; these are defined by the condition
\begin{equation}
F_{g,\alpha }^{\prime }(u)=\alpha +g^{\prime }(u)-s^{\prime }(u)=0,
\end{equation}
assuming that $s(u)$ and $g(u)$ are differentiable functions of $u$.
Assuming further that these functions are twice differentiable, we can take
the second derivative of $F_{g,\alpha }(u)$ with respect to $u$ to obtain
\begin{equation}
F_{g,\alpha }^{\prime \prime }(u)=g^{\prime \prime }(u)-s^{\prime \prime
}(u).
\end{equation}
From this equation, we readily see that the stability of the critical points
of $F_\beta (u)=F_{g=0,\alpha =\beta }(u)$ can be changed at will by
choosing various functions $g$. Suppose, for example, that $\hat{u}$ is
local maximum of $F_\beta (u)$ satisfying $s^{\prime }(\hat{u})=\beta $ and
$s^{\prime \prime }(\hat{u})>0$. Selecting $g$ such that $g^{\prime \prime }(
\hat{u})>s^{\prime \prime }(\hat{u})$, and choosing $\alpha $ in the
generalized canonical ensemble to be equal to the difference $s^{\prime }(
\hat{u})-g^{\prime }(\hat{u})$, we have now that $\hat{u}$ must be a
minimizer of $F_{g,\alpha }(u)$, for then we have $F_{g,\alpha }^{\prime }(
\hat{u})=0$, and $F_{g,\alpha }^{\prime \prime }(\hat{u})>0$ instead of
$F_\beta ^{\prime \prime }(\hat{u})<0$. This shows that a maximizer of
$F_\beta (u)$ can be changed to a minimizer of $F_{g,\alpha }(u)$.

Whether or not $\hat{u}$ is a local or a global minimizer of $F_{g,\alpha
}(u)$ is undetermined at this point. To distinguish between the two types of
minimizers, we need further notions of convex analysis that we shall not
discuss here; see Ref.~\cite{ellis2000}. However, it is possible to see
with a specific example that the critical points of $F_\beta (u)$ can be
transformed into global minima of $F_{g,\alpha }(u)$ without any problems.
Indeed, consider a quadratic function $g$ of the form $g(u)=\gamma (u-\hat{u}
)^2$, where $\hat{u}$ is a critical point of $F_\beta (u)$. The function
$F_{g,\alpha }(u)$ of the generalized canonical ensemble has then the form
\begin{equation}
F_{\gamma ,\beta }(u)=\alpha u+\gamma (u-\hat{u})^2-s(u).
\end{equation}
Its first derivative evaluated at $\hat{u}$ is
\begin{equation}
F_{\gamma ,\beta }^{\prime }(\hat{u})=\alpha -s^{\prime }(\hat{u})=0,
\end{equation}
so that the critical point $\hat{u}$ of $F_\beta (u)$ is also a critical
point of $F_{\gamma ,\alpha =\beta }(u)$. Now to be sure that this critical
point of $F_{\gamma ,\beta }(u)$ is a global and not just a local minimum of
$F_{\gamma ,\beta }(u)$, we only have to choose a large, positive value for
$\gamma $ that will make the term $\gamma (u-\hat{u})^2$ dominate in
$F_{\gamma ,\beta }(u)$. Changing $\gamma $ in this case does not change the
fact that $\hat{u}$ is a critical point of $F_{\gamma ,\alpha }(u)$, but it
changes its ``height'' with respect to the other critical points of
$F_{\gamma ,\alpha }(u)$.

This transformation of the critical points of some function to global minima
of a different function involving an added quadratic term is well-known in
the field of optimization \cite{bertsekas1982}, and has been described in
the specific context of nonequivalent ensembles by Ellis, Haven and
Turkington \cite{ellis2002}. Our work on the generalized canonical ensemble
can be seen as an extension of this previous work: it generalizes the
quadratic-transformation trick to arbitrary penalty functions $g$ in the way
sketched above. At the statistical-mechanical level, the added function $g$
leads, as we have seen, to the definition of a generalized canonical
ensemble, and the basic property that $g$ must possess in order to ensure
that local minima or local maxima of $F_\beta (u)$ are properly transformed
into global minima of $F_{g,\alpha }(u)$ can be related physically to the
existence of first-order transitions in the generalized canonical ensemble
\cite{costeniuc2006}.

To understand this last point, recall that the metastable and unstable mean
energies of $F_{g,\alpha }(u)$ lie in a forbidden range of mean-energy
values for $u_\beta $, whose existence defines physically a first-order
phase transition in the canonical ensemble. By transforming these metastable
and unstable mean energies of the canonical ensemble to stable mean energies
of a generalized ensemble, what one does, in effect, is to ``shorten'' or
``inhibit'' the range of forbidden mean energies in the generalized
ensemble. That is, by choosing $g$, one seeks to make the jump of $u_\beta $
seen in the canonical ensemble (Fig.~\ref{metae1}c) disappear in the
generalized canonical ensemble. With this in mind, one can guess that the
ultimate choice of $g$ is one that makes the function $s(u)-g(u)$ concave.
In this case, $F_{g,\alpha }(u)$ must be concave for any value $\alpha $,
which implies that $F_{g,\alpha }(u)$ can possess only one global minimizer
for every $\alpha $. Local minimizers or local maximizers are then not
allowed, whereby it can be proved that one recovers full equivalence between
the microcanonical and generalized canonical ensemble. The complete details
as to which choices of $g$ produce this effect can be found in
Refs.~\cite{costeniuc22005,costeniuc2006}; see, e.g., Theorems 3 and 4 in
Ref.~\cite{costeniuc2006}, as well as Fig.~3 in that reference.

To conclude this paper, we comment on four points:

(i) Our discussion of metastability was confined to the level of the
energy, but it can be generalized to describe metastability at the
macrostate level as well. Many spin models are known, for example, to have
magnetization states which are seen at equilibrium in the microcanonical
ensemble as a function of the energy, but not in the canonical ensemble as
function of the temperature; see, e.g., Refs.~\cite{antoni2002,ellis2004,costeniuc2005}.
Such nonequivalent magnetization
states correspond to metastable or unstable magnetization states in the
canonical ensemble, and could, in theory, be transformed to equilibrium,
stable states of a properly chosen generalized canonical ensemble. Work on
this topic is ongoing \cite{ellis2005}.

(ii) The generalized canonical ensemble ($g\neq 0$) can be interpreted
physically as describing a system in thermal contact with a
\textit{finite-size} heat bath, as opposed to the canonical ensemble ($g=0$) which
describes systems in thermal contact with an infinite-size heat bath. For
more details on this interpretation, the reader is referred to the original
work of Challa and Hetherington which introduced the Gaussian ensemble
\cite{challa21988,challa1988}, as well as Refs.~\cite{reif1965,johal2003}.

(iii) The generalized canonical ensemble describes extensive systems. In
defining it, it is assumed that the Hamiltonian $H$ scales like $O(n)$ in
the thermodynamic limit, that $\ln Z_g(\alpha )$ scales also like $O(n)$,
and that the limit defining the microcanonical entropy in Eq.(\ref{ms1})
exists.

(iv) Toral recently showed by direct calculation that the equilibrium
properties of a mean-field spin model can be obtained by using a modified
form of the partition function, which can be thought of to define a
generalized canonical ensemble \cite{toral2004}.
His generalization of the partition function
is related to our own generalization of the canonical ensemble (compare
Eq.(4) in Ref.~\cite{toral2004} with Eq.(\ref{gpf1}) of this paper).
Unfortunately, Toral does not
provide in his work any criteria for determining when the generalized
canonical ensemble is actually useful, that is, when it is equivalent to the
microcanonical ensemble or even the canonical ensemble. These criteria can
be found in Refs.~\cite{costeniuc22005,costeniuc2006}.

\begin{acknowledgments}
One of us (H.T.) would like to thank the organizing committee of the
NEXT-$\Sigma \Phi $ 2005 Conference for the invitation to give a talk. H.T. is
supported by NSERC\ (Canada), the Royal Society of London, and the School of
Mathematical Sciences at Queen Mary, University of London. The research of
R.S.E. is supported by the National Science Foundation (NSF-DMS-0202309).
B.T. is also supported by the National Science Foundation (NSF-DMS-0207064).
\end{acknowledgments}

\onecolumngrid

\end{document}